\documentclass[rnote]{aa} 
\usepackage{graphicx}
\usepackage{natbib}
\def\kappaasp{\kappa^{\rm (Asp)}}
\def\kappagn{\kappa^{\rm (GN)}}
\def\note #1]{{\bf #1]}}
\def\fig{.}
\begin{document}
   \title{On the opacity change required to compensate for
the revised solar composition}

\titlerunning{Opacity change compensating for revised solar composition}
\authorrunning{J. Christensen-Dalsgaard et al.}

   \author{J{\o}rgen Christensen-Dalsgaard
          \inst{1}
	  \and
	  Maria Pia Di Mauro
          \inst{2}
	  \and
	  G\"unter Houdek
          \inst{3}
	  \and
	  Frank Pijpers
          \inst{1}
          }

   \institute{Department of Physics and Astronomy, Building 1520,
	      University of Aarhus, DK-8000 Aarhus C, Denmark\\
              \email{jcd@phys.au.dk, fpp@phys.au.dk}
         \and INAF - IASF, Istituto di Astrofisica Spaziale e Fisica Cosmica,
            Via Fosso del Cavaliere 100, 00133 Roma, Italy\\
	    \email{mariapia.dimauro@iasf-roma.inaf.it}
         \and
	   Institute of Astronomy, Madingley Road, Cambridge CB3 0HA, UK
           \email{hg@ast.cam.ac.uk}
             }

   \date{Received XXXX; accepted XXXX}

 
  \abstract
   {Recent revisions of the determination of the solar composition have
    resulted in solar models in marked disagreement with helioseismic 
    inferences.}
   {The effect of the composition change on the model is largely caused by
    the change in the opacity. Thus we wish to determine an intrinsic opacity
    change that would compensate for the revision of the composition.}
   {By comparing models computed with the old and revised composition we
    determine the required opacity change.
    Models are computed with the opacity thus modified and used as reference
    in helioseismic inversions to determine the difference between the solar 
    and model sound speed.}
   {An opacity increase varying from around 30 per cent near the base of 
   the convection zone to a few percent in the solar core results in a 
   sound-speed profile, with the revised composition, which is
   essentially indistinguishable from the original solar model.
   As a function of the logarithm of temperature this is well represented by
   a simple cubic fit.
   The physical realism of such a change remains debatable, however.}
   {}


   \maketitle
%

\section{Introduction}

%
The opacity in the solar interior, and hence the solar internal structure,
depends sensitively on the abundances of the heavy elements
\citep[e.g.,][]{Turck2001a, Turck2008}.
Recent analyses of the solar spectrum have led to substantial revisions
of the solar abundances, particularly of oxygen, carbon and nitrogen
\citep[for a review, see][]{Asplun2005}.
Relative to previous work these studies have the advantage of being based
on three-dimensional hydrodynamical models of the solar atmosphere and
taking departures from local thermodynamic equilibrium into account.
Also, unlike earlier analyses they result in consistent abundance
determinations from different spectral lines.
As a result of the revision, the ratio $Z_{\rm s}/X_{\rm s}$ between 
the present solar surface abundances by mass of heavy elements and hydrogen is 
determined to be 0.0165, corresponding, in calibrated solar models,
to $Z_{\rm s} = 0.0125$.
For comparison, the commonly used composition of \citet{Greves1993}
yields $Z_{\rm s}/X_{\rm s} = 0.0245$, resulting in $Z_{\rm s} = 0.0181$.

   \begin{figure}
   \centering
   \includegraphics[width=8cm]{\fig/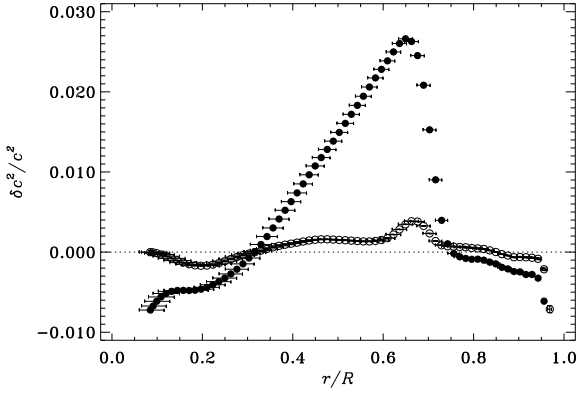}
      \caption{Inferred relative differences in squared sound speed,
      in the sense
      (Sun) $-$ (model), from inversion of the `Best Set' of observed
      frequencies of \citet{Basu1997}.
      The open circles show results for Model~S of \citet{Christ1996},
      using the old solar composition, while the filled circles show results
      for a corresponding model, assuming the revised composition.
      The horizontal bars indicate the resolution of the inversion while the
      vertical bars (hardly visible on this scale) show the $1{-}\sigma$ 
      errors in the inferences.
              }
         \label{fig:aspinv}
   \end{figure}

As pointed out, e.g., by 
\citet{Basu2004}, \citet{Montal2004}, \citet{Turck2004}, \citet{Bahcal2005} and 
\citet{Delaha2006}
this revision has substantial effects on solar models, greatly increasing 
the difference between their internal sound speed and the sound speed
inferred from helioseismology.
As an example, we consider Model~S
of \citet{Christ1996}, using the Grevesse \& Noels composition
and OPAL opacities from \citet{Iglesi1992}, and
a corresponding model based on the new abundances
and updated OPAL opacities \citep{Iglesi1996}.
Fig.~\ref{fig:aspinv} shows sound-speed differences between the Sun and
these two models, inferred through
inversion of the `Best Set' of observed frequencies of
\citet{Basu1997}, combining data obtained with the BISON network
and the LOWL instrument 
\citep[for further details on the inversion, see][]{Christ2007}.
Large differences are also found between models based on the revised
composition and the helioseismically
inferred depth of the convection zone and envelope helium abundance.
\citet{Chapli2007} found that analysis of low-degree solar oscillations
strongly supported the old heavy-element abundance;
on the other hand, the results of the analysis by \citet{Houdek2007},
with careful inclusion of the influence of the outer layers of the Sun,
indicated a heavy-element abundance somewhat lower
than the Model S value, although substantially above the value obtained
with the revised abundances.
The effects on solar models of the new composition,
as tested with helioseismology, were reviewed by \citet{Basu2008}.
\citet{Guzik2006} provided an overview of the, largely unsuccessful, attempts to
modify the assumptions in the model computation to compensate for the
composition change.



By far the most important effect on solar modelling of the heavy-element
abundance arises through the opacity.
Thus \citet{Montal2004} found that a substantial opacity increase
near the base of the convection zone would help reducing the
discrepancy caused by the revised abundances.
Similarly, it was noted by \citet{Bahcal2005} that an intrinsic change in the
opacity could be used to correct the model computation, and they estimated
that an opacity increase of around 11 per cent over a relatively broad
range in temperature would be required.
Indeed, there are undoubtedly significant uncertainties in the very 
complex opacity calculations.
In the present note, following Bahcall et al.,
we make a more detailed analysis of this nature, estimating the
intrinsic change in the opacity required to obtain a model structure
corresponding largely to Model~S, but with the revised composition.

\section{Determination of the opacity change}

%
The goal is to determine an opacity modification such that the
sound-speed structure of Model~S can be approximately reproduced with the
revised solar surface composition.
With just this constraint we can clearly only determine a modification that
depends on a single variable which we take to be temperature $T$.
Thus we write the modified opacity as
\begin{equation}
\log \tilde \kappa(\rho, T, X, Z) =
\log \kappaasp(\rho, T, X, Z) + f(\log T) \; ,
\end{equation}
where $\rho$ is density, $X$ and $Z$ are the abundances by mass of hydrogen
and heavy elements, and $\kappaasp$ is the opacity evaluated with the
revised heavy-element composition;
$\log$ denotes logarithm to base 10.
The goal is to determine $f(\log T)$ such that $\tilde\kappa$ evaluated
for a structure corresponding to Model~S, but with the revised heavy-element
composition, matches the opacity $\kappagn$ in Model~S,
evaluated with the original \citet{Greves1993} heavy-element composition.

   \begin{figure}
   \centering
   \includegraphics[width=8cm]{\fig/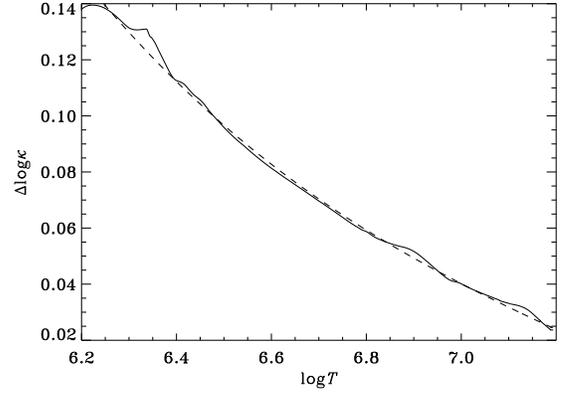}
      \caption{The solid curve shows the intrinsic opacity change
       $\Delta \log \kappa = f(\log T)$
       determined from Eq.~(\ref{eq:dlkap}), on the basis of models
       with the old and the revised composition.
       The dashed curve shows the cubic fit $f_{\rm approx}$ to
       $\Delta \log \kappa$, given by Eq.~(\ref{eq:fit}).
              }
         \label{fig:dlkap}
   \end{figure}


To estimate $f(\log T)$ we consider two models of the present Sun:
one is Model~S, characterized by $\{\rho_{\rm S}, T_{\rm S},
X_{\rm S}, Z_{\rm S}\}$, as functions of position in the model, and the
second is Model Asp, similarly characterized by
$\{\rho_{\rm Asp}, T_{\rm Asp}, X_{\rm Asp}, Z_{\rm Asp}\}$.
We determine the difference in $\log\kappa$, at fixed $T$, between the two
models as
\begin{eqnarray}
\delta_T\log \kappa
= && \log \kappagn(\rho_{\rm S}(T), T, X_{\rm S}(T), Z_{\rm S}(T)) \\
&& - \log \kappaasp(\rho_{\rm Asp}(T), T, X_{\rm Asp}(T), Z_{\rm Asp}(T)) 
\nonumber \; ,
\end{eqnarray}
making explicit that the models are computed with $\kappagn$ and $\kappaasp$,
respectively.
Using the required property of $\tilde\kappa$, and linearizing in model
differences, we obtain
\begin{eqnarray}
\delta_T\log \kappa
\simeq && f(\log T) + 
\left( {\partial \log \kappa \over \partial \log \rho}\right)_{T, X} 
\delta_T\log \rho
\nonumber \\
&& + \left( {\partial \log \kappa \over \partial \log X}\right)_{T, \rho}
\delta_T\log X \; ,
\end{eqnarray}
neglecting the effect of the different dependence of $Z$ on position
in the two models;
here, e.g., $\delta_T \log \rho = \log\rho_{\rm S}(T) - \log\rho_{\rm Asp}(T)$.
A similar expression was obtained by \citet{Bahcal2005}.
Hence, neglecting the relatively modest effects of the
differences in $X$ and $Z$ at fixed $T$,
we obtain the required opacity change as
\begin{equation}
\Delta \log \kappa = f(\log T) \simeq \delta_T \log \kappa -
\left( {\partial \log \kappa \over \partial \log \rho} \right)_{T, X}
\delta_T \log \rho \; .
\label{eq:dlkap}
\end{equation}
This procedure can be iterated, to compensate for the error in the
linearization and the neglect of the composition effects.
In practice we have found that two iterations are sufficient to reach
a model closely matching Model~S.

   \begin{figure}
   \centering
   \includegraphics[width=8cm]{\fig/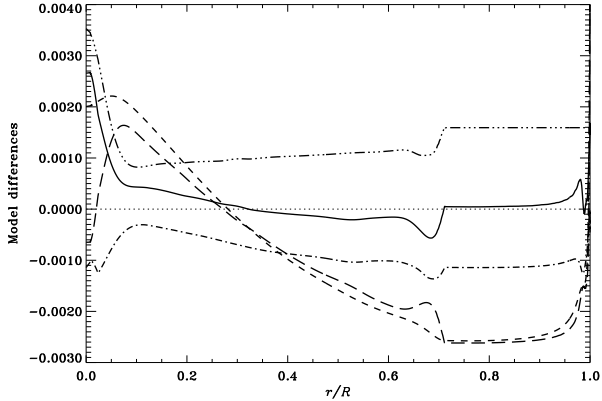}
      \caption{Differences between Model~S${}^\prime$, computed with the
revised composition and the opacity change $\Delta \log \kappa$ illustrated
in Fig.~\ref{fig:dlkap}, and Model~S, in the sense 
(Model~S${}^\prime$) $-$ (Model~S):
$\delta \ln c^2$ (continuous), 
$\delta\ln p$ (short dashed),
$\delta\ln \rho$ (long dashed),
$\delta\ln T$ (dot-dashed) and
$\delta X$ (triple-dot-dashed). 
Here $\ln$ is natural logarithm.
              }
         \label{fig:moddif}
   \end{figure}


\section{Results}

\begin{figure}
   \centering
   \includegraphics[width=8cm]{\fig/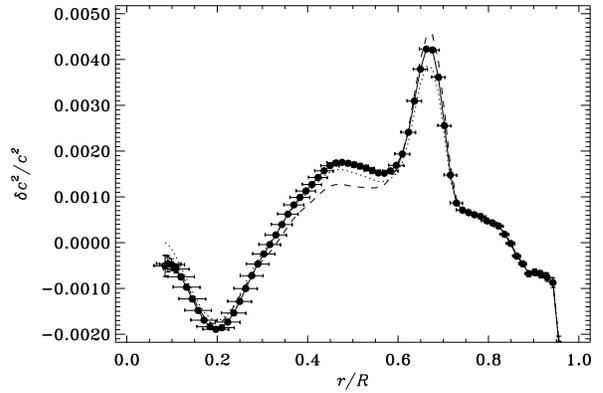}
      \caption{Inferred relative differences in squared sound speed, 
in the sense
(Sun) $-$ (model), from inversion of the `Best Set' of \citet{Basu1997}.
The symbols with horizontal and vertical bars (see caption to 
Fig.~\ref{fig:aspinv}) show results for Model~S${}^\prime$, computed with
the revised composition and applying the opacity correction 
$\Delta \log \kappa$ shown in Fig.~\ref{fig:dlkap}.
The dashed curve shows differences for \hbox{Model~S${}^{\prime\prime}$},
similarly computed using the fit in Eq.~(\ref{eq:fit}).
For comparison, the dotted curve shows the inferred difference for the
original Model~S.
              }
         \label{fig:modinv}
   \end{figure}

%
Figure~\ref{fig:dlkap} shows the resulting opacity change, largely restricted to
the radiative interior which evidently is the only region where the
change is relevant. 
We have computed a full evolution sequence assuming the revised surface
composition and applying this change to the opacity,
and, as for Model~S, calibrating the model to solar luminosity and
radius as well as to the revised $Z_{\rm s}/X_{\rm s}$.
Differences between the resulting Model~S${}^\prime$
and Model~S are illustrated in Fig.~\ref{fig:moddif}.
It is evident that the model matches Model~S very closely. 
To test the effect on the comparison with the helioseismic inference,
Fig.~\ref{fig:modinv} shows the sound-speed inversion
using this model as reference.
Clearly it matches the helioseismic results as well as does Model~S.

\begin{table}
\caption{Properties of solar models (see text for a description).
$D_{\rm cz}$ is the depth 
of the convective envelope, given in units of the solar radius ${\rm R}_\odot$,
and $Y_{\rm e}$ is the helium abundance in the envelope.}
\label{table:1}      
\centering                          
\begin{tabular}{l c c}        
\hline\hline                 
\noalign{\vskip 3pt}
Model & $D_{\rm cz}/{\rm R}_\odot$ & $Y_{\rm e}$ \\
\hline                        
\noalign{\vskip 3pt}
   Asp                  & 0.2712 & 0.2286 \\
   S                    & 0.2885 & 0.2447 \\
   S${}^\prime$         & 0.2882 & 0.2489 \\
   S${}^{\prime\prime}$ & 0.2875 & 0.2488 \\
\hline                                   
\end{tabular}
\end{table}

The simple dependence of $f(\log T)$ on $\log T$ makes it natural to
approximate it by a low-order polynomial.
In Fig.~\ref{fig:dlkap} the dashed line shows the following fit:
\begin{equation}
f_{\rm approx}(\log T) = 0.1298 - 0.1856 \xi + 0.1064 \xi^2 - 0.0345 \xi^3 \; ,
\label{eq:fit}
\end{equation}
with $\xi = \log T - 6.3$,
obtained with a least-squares fit with uniform weight to $f(\log T)$,
for $\log T > 6.3$.
The result of using the modification in the evolution calculation,
calibrating the model as before, and using the resulting
\hbox{Model~S${}^{\prime\prime}$}
as reference in a sound-speed inversion is also shown in Fig.~\ref{fig:modinv}.
This is barely distinguishable from the result of using the original
$f(\log T)$.

We finally list in Table~1 the values of the depth $D_{\rm cz}$
of the convection zone and the envelope helium abundance $Y_{\rm e}$
for the models considered.
For comparison, the helioseismically inferred value of $D_{\rm cz}$ is
around $0.287$ and $Y_{\rm e}$ has been determined to be around 
$0.25$, although with some sensitivity to the equation of state used in the
solar modelling
\citep[for a review, see][]{Basu2008}.
It is evident that Model~Asp is inconsistent with the observed values, 
whereas the remaining models are essentially in accordance with
observations.

\section{Discussion and conclusion}

%
The change in $\log \kappa$ obtained in Fig.~\ref{fig:dlkap}
corresponds to an opacity change of around 30 per cent
near the base of the solar convection zone,
decreasing to a very modest level in the core.
This immediately raises the question whether such a change is physically
realistic.
Comparisons between independent opacity calculations
\citep[e.g.,][]{Badnel2005} \citep[see also][for a review]{Basu2008}
indicate that the precision of the opacities in the relevant temperature
range is better than 5 per cent, far smaller than the required change.
On the other hand it seems possible that, although highly sophisticated,
the present opacity calculation might neglect significant effects.
Kurucz (personal communication) has noted that the neglect of 
a large number of elements of low abundance could have a significant
effect on the Rosseland mean opacity, which is sensitive to even rather
weak absorption in spectral bands not affected by lines of the more
abundant elements.


It is perhaps relevant to recall the somewhat similar situation more than
two decades ago when \citet{Simon1982} made a plea for the 
reexamination of opacity calculations in the light of problems with the
modelling of certain pulsating stars;
he suggested that an increase in the opacity by a substantial factor
could remove the discrepancies between models and observations.
Although \citet{Magee1984} claimed that such an increase would be
`incompatible with atomic physics'
it was in fact found in the OPAL opacity calculations
\citep[e.g.,][]{Iglesi1991, Rogers1992},
as a result of the inclusion of the effects of a large number of lines.
Based on this experience one should perhaps be wary of excluding the
possibility of substantial opacity modifications.


It is interesting that already the early helioseismic sound-speed
inferences by \citet{Christ1985} indicated a need for an increase 
in the then current opacities,
as subsequently confirmed by the OPAL calculations.
More recently, \citet{Tripat1998a} determined the opacity correction
to the OPAL tables required to match the difference between the 
helioseismically inferred sound speed and the sound speed in Model S,
using the sensitivity of sound speed to opacity changes determined by
\citet{Tripat1998b}; 
they found that a relative difference in opacity of order 5 per cent was
needed.
Also, \citet{Gough2004} determined the opacity difference between Model S and
a helioseismically calibrated solar model,
finding a relative difference of order 1.5 per cent.

We have assumed that the opacity correction is a function of temperature
alone.
This is obviously a gross, if unavoidable, simplification which should
be kept in mind if the correction obtained here is used for other
stellar-model calculations.
Also, we emphasize that the fit given in Eq.~(\ref{eq:fit}) is
only valid in the range $[6.3, 7.2]$ in $\log T$ over which it was
obtained.
Even so, it might be interesting to use an opacity change similar to the
one obtained here in stellar computations, such as the isochrone
analysis of M67 presented by \citet{Vanden2007}.


Given the difficulties arising for solar modelling from the revised
solar abundances it is obviously crucial to carry out further 
tests of the results.
Also, the properties of the computed atmosphere models should evidently
be tested against other relevant observations
\citep[e.g.,][]{Ayres2006}.
The complexity of the hydrodynamical modelling of the solar atmosphere
makes independent calculations highly desirable.
Thus it is encouraging that such calculations are now under way
\citep{Steffe2007, Caffau2007}.
Very recently, \citet{Caffau2008} made a determination of the
solar oxygen abundance, resulting in a value intermediate between the
old and new compositions considered here.
We expect that the corresponding opacity correction required to
match Model S, obtained as in the
present analysis, would be roughly half the value shown in Fig.~\ref{fig:dlkap}.
{Interestingly, \citet{Holweg2001} obtained an oxygen abundance consistent
with the value found by Caffau et al.;
\citet{Turck2004} showed that this did in fact result in
a sound-speed difference relative to the
helioseismic inferences intermediate between the results 
for the old and new compositions considered here.}

Alternative independent determinations of the abundances of the
relevant elements would obviously be very valuable.
An interesting possibility, proposed by
\citet{Gong2001} and reviewed by \citet{Basu2008},
is to constrain the heavy-element abundance from
helioseismic inference of its effect 
on the thermodynamic state, and hence the sound speed, in
the solar convection zone.
This is an extension of the successful helioseismic
determination of the solar-envelope helium abundance 
\citep[e.g.,][]{Voront1991, Kosovi1992, Antia1994},
following the suggestion of \citet{Gough1984} and \citet{Dappen1986}.
So far the results of such analyses are somewhat uncertain:
\citet{Lin2005} found slight indications 
that the standard \citep{Greves1993} heavy-element abundance
was too high, whereas \citet{Antia2006} and \citet{Lin2007}
concluded that the helioseismic results confirmed the original abundances.
Further analysis, investigations of the effects of the
uncertainties in the equation of state, as well as better data
including modes of higher degree, are required to obtain more definitive
results.

The relatively limited goal of the present note is to investigate
the compensating changes to the opacity required by the revision in the
solar heavy-element composition, in order to match models computed with
the earlier abundance determinations.
This illustrates one aspect of the sensitivity of the solar structure,
as probed by helioseismology, to the physics of the solar interior.
It is evident that a broader goal of helioseismology is to understand the
full range of solar internal microphysics and dynamics required to obtain a
model in accordance with the helioseismic inferences.
As discussed above, this may involve further adjustments of the opacity;
however, it is likely that other processes, such as weak mixing in the
region below the convection zone, must be invoked
\citep[e.g.,][]{Brun1999, Christ2007}.
Through such suitable adjustments to the physics used in the modelling
it is possible to construct seismic solar models that match 
the inferred sound speed \citep[e.g.,][]{Turck2001b};
a more interesting question is evidently the physical basis for these
adjustments.

\begin{acknowledgements}
We are grateful to D. O. Gough, R. L. Kurucz and A. Weiss
for useful conversations.
This work was supported by the European Helio- and Asteroseismology Network 
(HELAS), a major international collaboration funded by the 
European Commission's Sixth Framework Programme.
\end{acknowledgements}

\end{document}